\documentclass{aa}
\usepackage{graphicx}
\usepackage{txfonts}
\begin{document}
   \title{Application of the Trend Filtering Algorithm 
          on the MACHO Database}

   \author{Szul\'agyi, J.\inst{1,2}, Kov\'acs, G.\inst{2}, Welch, D.~L.\inst{3}}

%   \offprints{\email{kovacs@konkoly.hu}}

   \institute{
   E\"otv\"os University, Department of Astronomy, 
   Budapest, P.O. Box 32, H-1518 Hungary \\ 
   \email {szulagyi@konkoly.hu}
   \and
   Konkoly Observatory, P.O. Box 67, H-1525,
   Budapest, Hungary \\ 
   \email {kovacs@konkoly.hu} 
   \and
   Department of Physics and Astronomy, McMaster University,
   Hamilton, Ontario, Canada, L8S 4M1 \\ 
   \email {welch@physics.mcmaster.ca}
   }

   \date{Received / Accepted }

%
%%%%%%%%%%%%%
% Abstract  %
%%%%%%%%%%%%%
%
   \titlerunning {MACHO data trend filtering}
   \abstract
% context heading (optional), leave it empty if necessary
{}
% aims heading (mandatory)
{Due to the strong effect of systematics/trends in variable star 
observations, we employ the Trend Filtering Algorithm (TFA) on a 
subset of the MACHO database and search for variable stars.}
% methods heading (mandatory)
{TFA has been applied successfully in planetary transit searches, 
where weak, short-lasting periodic dimmings are sought
in the presence of noise and various systematics (due to, e.g., 
imperfect flat fielding, crowding, etc). These latter effects 
introduce colored noise in the photometric time series that can 
lead to a complete miss of the signal. By using a large number of 
available photometric time series of a given field, TFA utilizes 
the fact that the same types of systematics appear in several/many 
time series of the same field. As a result, we fit each target time 
series by a (least-square-sense) optimum linear combination of 
templates and frequency-analyze the residuals. Once a signal is 
found, we reconstruct the signal by employing the full model, 
including the signal, systematics and noise.} 
% results heading (mandatory)
{We apply TFA on the brightest $\sim$5300 objects from subsets of 
each of the MACHO Large Magellanic Cloud fields \#1 and \#79. We find 
that the Fourier frequency analysis performed on the original data 
detect some 60\% of the objects 
as trend-dominated. This figure decreases 
essentially to zero after using TFA. Altogether, We detect 387 variables 
in the two fields, 183 of which would have remained undetected without 
using TFA. Where possible, we give preliminary classification of the 
variables found.} 
% conclusions heading (optional), leave it empty if necessary
{}


   \maketitle
%

%%%%%%%%%%%%%%%%%
%  SECTION 1
%%%%%%%%%%%%%%%%%
%
\section{Introduction}
Microlensing survey projects initiated in the 1990s yielded unprecedented 
amounts of data for variable star research. Subsequent analyses and 
followup observations focused on certain types of variables and largely 
ignored the several millions additional lower-amplitude variable stars 
spanning the HR-diagram, including those not suspected of belonging to 
existing classes of stellar variability. Since then, a new class of 
intensive photometric survey has emerged motivated by the search for 
transiting extrasolar planets. Here, due to the unexpected nature of 
various extrasolar planets, the analysis is not restricted to a subsample 
of the collected time series. It quickly became evident from these 
surveys that the expected faint signals are usually overwhelmed by the 
much larger systematic effects (or trends) due to imperfect conditions, 
hardware and photometric data reduction. Until recently, image and 
photometric reduction methods were unable to retrieve the bulk of 
variability information in these collections of data. To cure this 
deficiency, methods have been developed such as the one of Kruszewski 
\& Semeniuk (2003), SysRem by Tamuz, Mazeh \& Zucker (2005) and TFA 
by Kov\'acs, Bakos, \& Noyes (2005, hereafter KBN) to filter out 
systematics in a post-processing phase. The method of 
Kruszewski \& Semeniuk (2003) is devoted to reconstruct the signal 
with known period. This includes a combined fit of a Fourier series 
(with the signal period) and a polynomial with the hour angle. On the 
other hand, SysRem and TFA are capable of searching for and reconstructing 
signals by using the temporal characteristics of the objects observed 
in the field to find the part of the target time series that comes 
from the systematics.  

Attempts have been made to deal with the most easily identified 
systematics and to correct for them. For example, the method of 
External Parameter Decorrelation (EPD, see Bakos et al. 2009) utilizes 
the fact that, for constant stars, certain image properties (e.g., 
position, PSF width and elongation) correlate with the brightness 
deviations from the average value. There is also the method of 
Differential Image Analysis (DIA, see Alard \& Lupton 1998 and for a 
summary Bramich 2008) that has been used very successfully in variable 
star searches over the past ten years. None of these methods result in 
photometric time series that are completely free of systematics. This 
is because they use only the spatial information on the images and 
do not include the temporal information, which is available in great 
abundance from long-term observations. 

We briefly describe our method in Sect.~2. In Sect.~3, for the 
preparation of the data analysis, we perform various tests to optimize 
our search for variables in Sect.~4. Discussion and summary of the 
results are given in Sect.~5.      
 
%%%%%%%%%%%%%%%%%
%  SECTION 2
%%%%%%%%%%%%%%%%%
%
\section{The trend filtering method}
The various aspects of the trend filtering method has already been 
described in several other papers (e.g., KBN, Kov\'acs \& Bakos 2008). 
Here we briefly summarize the basic steps and encourage the reader 
to consult with those papers for more details. 

The basic idea behind TFA is that any systematics (or trends -- we 
use these terms interchangeably) contaminating the target signal can 
be linearly decomposed with the aid of some subset of the photometric 
signals (or light curves, LCs for simplicity) of the objects observed 
in the given field. This template lightcurve set is expected to show 
the various types of brightness variations which are common in many 
stars (constant or variable) in the field. There are two steps in the 
analysis:
\begin{itemize}
\item
Assuming that the target signal is trend- and noise-dominated, 
we fit the target signal by the linear combination of a pre-selected 
template LC set. This fit yields a residual time series that is 
supposedly trend-free and contains only the signal (with some 
distortion -- see below) and noise, that is nearly white (compared 
to the starting noise spectrum). This filtered time series is 
frequency-analyzed and searched for dominant frequency components.   
\item 
Once the frequencies are found, one may want to {\em restore} the 
signal, since in the above procedure we assumed that there are only 
systematics and noise. Since some part of the original signal may 
have been fitted and subsequently subtracted (unintentionally) from 
the original time series, we need to restore the signal by employing 
a {\em full} model that contains the {\em signal+trend+noise}. Once 
the constituting frequency components are known, we can perform this 
reconstruction by iteratively approximating the signal and trend part 
of the observed signal. For periodic signals the method has been 
described in KBN, for multiperiodic ones we refer to Kov\'acs \& Bakos 
(2008). 
\end{itemize}

We note the following. The template set is selected in a nearly 
uniform $(X,Y)$ grid covering the full field. This distribution 
of templates makes the selection of close pairs fairly unlikely, 
thereby decreasing the chance of eliminating blended variables 
that could be further studied by other means (e.g., by large-telescope 
followup observations). A lower brightness limit is also set, 
since very faint targets are dominated mostly by random noise 
and therefore are less valuable for trend filtering (see however 
Mazeh, Tamuz \& Zucker 2007 for possible systematics in faint stars). 
We do not omit variables from the template, because: (i) variables 
are relatively rare (e.g., some 5\% of the Galactic field stars 
are variable -- see Kov\'acs \& Bakos 2008); (ii) we do not know 
{\em a priori} which stars are variable and; (iii) we do not generate 
additional signal in the target time series by including variables 
in the template set (see KBN for the discussion of the effect of a 
template with sinusoidal variation on a target of pure noise). 

For time series of limited number of data points, large numbers of 
template (comparable with the number of data points) will generate 
correlated noise in the filtered time series. Although this is a 
general feature of any type of data fitting, the effect is still 
likely be tolerable when compared with the level of correlation 
caused by the systematics in the original time series. To avoid 
over-fitting, we compute the unbiased estimate of the standard 
deviation of the residuals and select a template number for which 
this standard deviation starts to level off. As an additional test 
we also show the insufficiency of the use of a single, but optimally 
selected template as a substitute of multiple templates (see Sect.~3.2).    

The multiple spatial template requirement is consistent with the 
manner in which the photometry in the MACHO Project database was 
originally reduced. Alcock et al (1999) describe reduction and 
calibration of the MACHO Project photometry, noting that each image 
was broken into $512\times512$ pixel subimages for separate, 
parallel-efficient reduction. Furthermore, a focal plane-wide 
pattern of zero point residuals was found, likely due to 
variations in the large, dichroic beamsplitter.

%%%%%%%%%%%%%%%%%
%  SECTION 3
%%%%%%%%%%%%%%%%%
%
\section{Preparation of the analysis}
Before applying TFA on MACHO photometry, it is important to 
investigate the basic properties of the data, because we would 
like to use the method in an optimum way, by filtering out 
systematics at the lowest possible filter order (i.e., template 
number $N_{\rm TFA}$).  

Because our earlier experiments with the 
HATNet\footnote{{\bf H}ungarian-made {\bf A}utomated 
{\bf T}elescope {\bf Net}work, see 
{\em http://cfa-www.harvard.edu/~gbakos/HAT/}} data show that 
fairly large number of templates (in the order of several hundred, 
see KBN) are necessary to reach the desired white noise level, we 
needed to investigate this issue in the particular case of MACHO 
time series photometry in order to avoid over-fitting the data. 

In Sect.~3.1 we describe the data used in this paper. Sect.~3.2 is 
devoted to several tests, including those of finding the optimum 
template number, demonstrating the improvement in the distribution 
of the frequencies of the main harmonic components of the TFAd time 
series and showing the inadequacy of best single template fit. We 
also give an estimate on the false alarm level based on Gaussian 
time series simulations.

%%%%%%%%%%%%%%%%%
% Sunsection 3.1
%%%%%%%%%%%%%%%%%
%
\subsection{The data}
We selected sub-samples from two neighboring LMC fields (\#1 and \#79) 
to create sample databases containing some 5300 stars from the bright 
tail of the magnitude distribution of each field. Field \#79 is densely 
populated at the Northern edge of the LMC bar, containing altogether 
$532264$ objects. In practice, a MACHO Project ``object'' is a defined
location in the focal plane where a sum of a small number of images 
with better-than average image quality exceed the threshold of a the 
star-finding algorithm. An object may constitute a single star but also 
may be the combined light of several stars in a single MACHO Project 
resolving element.

Field \#1 overlaps \#79 and is situated further west at the end 
of the bar of the LMC. This is a somewhat sparser field with $503227$ 
objects detected by the MACHO survey. Both datasets cover a long and 
continuous time interval of $7.5$~years, starting in 1992. We use the 
MACHO ``b'' instrumental magnitudes, because of the better noise 
characteristics based on our earlier works on RR~Lyrae stars (Alcock 
et al. 2003).

To check the unbiased nature of our samples in terms of colors (and 
implicitly physical properties), we plot the color-magnitude diagrams (CMDs)  
in Fig.~1. To limit the sample size and exclude too many noisy targets but 
at the same time to reach the overall luminosity level of the RR~Lyrae 
stars, we applied a cut at $V=20$~mag. 

The transformation of the instrumental magnitudes to the (Cousins) 
$R$ and (Johnson) $V$ magnitudes have been discussed by Alcock et al. 
(1999). Here we use the formulae given by Alcock et al. (1997) for the LMC 
(their Eq. (1)). The figure shows that the sample used in our analysis 
covers the expected full color range of objects observed by the project 
(see the CMD of Alves et al. 1999 for nine million objects from the LMC). 

Because TFA requires exactly the same set of observation epochs for all 
LCs, it is important to check how well this uniform sampling condition 
is satisfied. Figure~2 shows the distribution of the number of data points 
on each LC. As expected, \#79 displays a different distribution, because 
of the more frequent observations of the LMC bar field. It is important 
to note that 80\% of the objects have number of photometric measurements 
in the ranges of 1180--1250 and 1460--1550 for \#1 and \#79, respectively. 
All stars in the TFA template sets are brighter than $V=18.6$~mag and none 
of the objects in either fields in the above samples have fewer than 1000 
data points (objects containing fewer data points were not selected -- 
we see from the figures that this cut is justified). This allows the 
application of TFA on all stars, with some special attention to be paid 
to the low-N cases, when the empty values are to be filled in by the 
average of the time series. We note that this necessary extension of 
the target time series has little effect on the final result, because: 
(i) there are only a few stars with significantly fewer data points than 
the average; (ii) we use a common sampling time base only for frequency 
search; (iii) the final result (i.e., the reconstructed LC) is based on 
``target-adapted'' data points, i.e., we sample the templates on the 
time base of the target. 

It is also useful to check the distribution of the standard deviations, 
since selection of outliers and other processes affecting the final data 
quality depend on this quantity. Figure~3 shows that both fields have 
similar error distributions, perhaps with a slight surplus of objects 
with errors larger than $\sim 0.2$~mag in field \#79. This suspected 
difference can be attributed to the $\sim 6$\% higher surface density 
of objects in \#79 relative to \#1. We note that in the analyses 
presented in this paper we employ a fairly liberal upper limit of 
$0.4$~mag for selecting erroneous data points based on the errors 
derived from the images by the MACHO pipeline. In addition, further 
data points are discarded by an iterative $5\sigma$-clipping based on 
the standard deviation of the time series. The overall value of the 
standard deviations of the TFA-filtered LCs toward the faint limit 
of $V=19.0$~mag is about $0.15$~mag. In assessing the minimum detectable 
amplitude in this regime, we assume that the stars are mostly dominated 
by white noise. From the approximate $A/(\sigma\sqrt{2/N})>7$ condition 
(valid in the high SNR limit -- see, e.g., Kov\'acs 1980), for field 
\#79 with an average number of data points of $N=1400$ we get a 
$7\sigma$ lower limit of $\sim 0.04$~mag for the amplitudes of periodic 
signals detectable even toward this faint end of our sample.

%>>>>>>>>>>>>>>>
%  FIGURE 1
%>>>>>>>>>>>>>>>
%
   \begin{figure}[h]
   \centering
   \includegraphics[width=90mm]{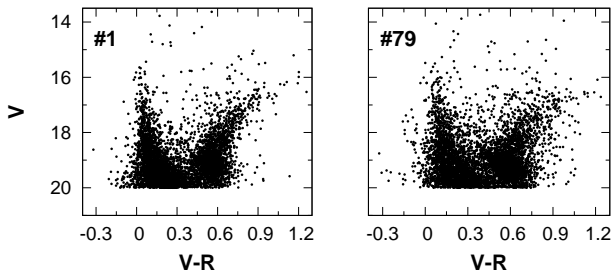}
      \caption{Color-magnitude diagrams for the stars analyzed in this 
               paper in the LMC fields \#1 and \#79. Data were cut at 
	       $V=20$~mag.} 
         \label{fig1}
   \end{figure}
%
%
%>>>>>>>>>>>>>>>
%  FIGURE 2
%>>>>>>>>>>>>>>>
%
   \begin{figure}[h]
   \centering
   \includegraphics[width=90mm]{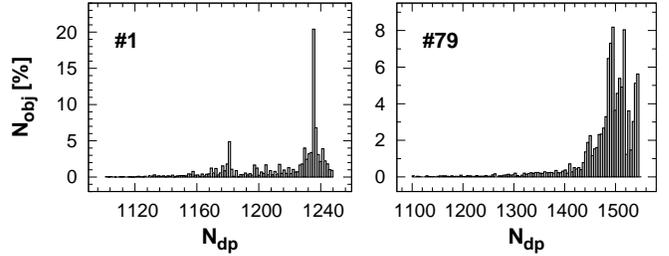}
      \caption{Distribution of the number of the constituting data points of 
               the light curves of the stars analyzed in this paper in the 
	       LMC fields \#1 and \#79.} 
         \label{fig2}
   \end{figure}
%
%
%>>>>>>>>>>>>>>>
%  FIGURE 3
%>>>>>>>>>>>>>>>
%
   \begin{figure}[h]
   \centering
   \includegraphics[width=90mm]{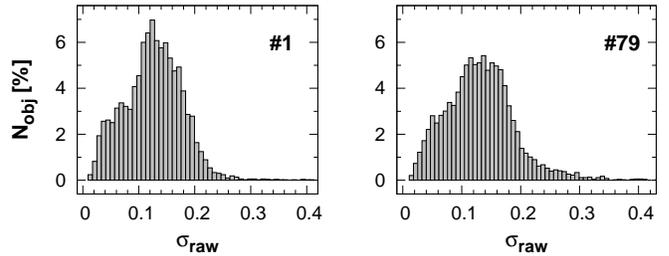}
      \caption{Distribution of the standard deviations of the light 
               curves analyzed in this paper from the LMC fields 
	       \#1 and \#79.} 
         \label{fig3}
   \end{figure}
%

%%%%%%%%%%%%%%%%%
% Subsection 3.2
%%%%%%%%%%%%%%%%%
%
\subsection{Tests for optimum analysis}
First we check the TFA template number ($N_{\rm TFA}$) dependence 
of $\sigma_{\rm TFA}$, the unbiased estimate of the standard deviation 
of the residuals remaining after subtracting the systematics from 
the original time series. Omitting the further decrease of the 
standard deviation due to signal reconstruction is reasonably 
justified, since less than 5\% of the stars are variables, where 
reconstruction really matters. The unbiased estimate is computed as 
given in Appendix A. 

Figure 4 shows the variation of $\sigma_{\rm TFA}$ for 100 randomly 
selected stars from each field, exhibiting integer d$^{-1}$ systematics. 
For comparison, we also plot the results obtained on pure Gaussian white 
noise data generated on the time base of the same stars. We see a clear 
decrease of $\sigma_{\rm TFA}$ with the increase of $N_{\rm TFA}$ for 
the observed data. The expected constant level of the test data is also 
well exhibited. From the individual results (thin lines) we see that 
there are fairly large differences among the various objects in respect 
of the run of $\sigma_{\rm TFA}$. For `well-behaved' systematics and 
extensive data sets we expect a steadily decreasing $\sigma_{\rm TFA}$ 
from low template numbers to some optimum value and then leveling off 
to a basically constant value. This pattern is approximately present 
only in a few cases. In general, we see a nearly monotonic decrease of 
$\sigma_{\rm TFA}$ all through the template numbers tested. We suspect 
that this behavior is related to the large variety of systematics that 
are still not possible to catch even with the largest number of 
templates tested. In any case, based on the above test and the one 
described below, as a compromise between overall data number and trend 
filtering level, we fix the template number to $600$ and $500$ for 
fields \#79 and \#1, respectively. 
 
%
%>>>>>>>>>>>>>>>
%  FIGURE 4
%>>>>>>>>>>>>>>>
%
   \begin{figure}[h]
   \centering
   \includegraphics[width=70mm]{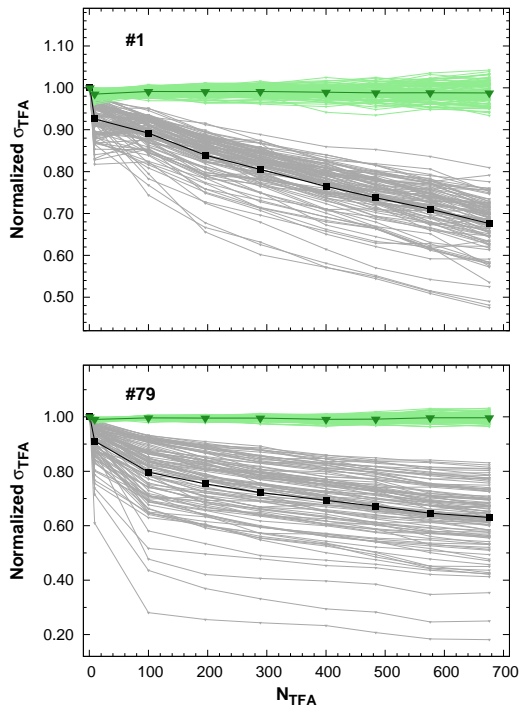}
      \caption{Dependence of $\sigma_{\rm TFA}$ (see Appendix A) on 
               $N_{\rm TFA}$. Near horizontal lines are for pure 
	       white noise test data. Thin lines are for the individual 
	       objects. Thick lines show the average values computed 
	       from the 100 objects. Each function is normalized to 1.0 
	       at $N_{\rm TFA}=0$, i.e., for the standard deviation of 
	       the original (non-TFAd) data.} 
         \label{fig4}
   \end{figure}

In a subsequent test we show that the above choice of template number 
is adequate for filtering out prominent systematics. We follow KBN, 
where the filtering capability was tested by the comparison of the 
occurrence rates of the peak frequencies found by the Box-fitting Least 
Squares (BLS) analysis (see Kov\'acs, Zucker \& Mazeh 2002) before 
and after the application of TFA. Here we perform the same type of 
test but we use Discrete Fourier Transformation (DFT, see e.g., Deeming 
1975), since we are mostly interested in the oscillation components 
of the data. The resulting distribution functions are shown in Figs.~5 
and 6. We see that the original (RAW) data exhibit very strong 
systematics associated with various forms of the common $1$d$^{-1}$ 
effects (e.g., change in the point spread function, color-dependent 
absorption, etc.). All of these are absent in the TFA-filtered data 
and the frequency distributions are nearly flat, as expected from a 
randomly selected sample of stars. At the low-frequency end, a part 
of the remaining objects are real (long-period) variables as we are 
to discuss them in Sect.~4. In addition to the cyclic systematics, 
TFA is capable to handle also outlier data points that originate from 
various transients (e.g., from non-properly subtracted cosmic rays). 
This property is obviously not exhibited in Figs.~5 and 6, but can be 
seen in several phase-folded light curves (in Sect.~4 we show an 
example on this).

%
%>>>>>>>>>>>>>>>
%  FIGURE 5
%>>>>>>>>>>>>>>>
%
   \begin{figure}[h]
   \centering
   \includegraphics[width=80mm]{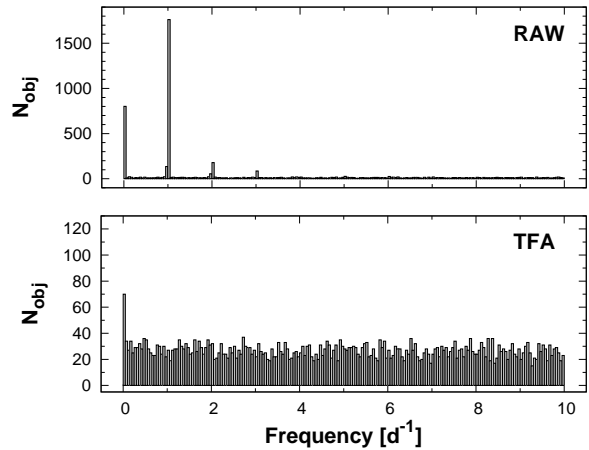}
      \caption{Distribution of the peak frequencies for the full sample 
               of $\sim$~5300 objects in field \#1. The DFT analysis was 
	       performed in the frequency range of $[0.0,10.0]$d$^{-1}$.
	       Upper panel: original (RAW, non-TFAd) data, lower panel: 
	       TFAd data with $N_{\rm TFA}=500$.} 
         \label{fig5}
   \end{figure}

%
%>>>>>>>>>>>>>>>
%  FIGURE 6
%>>>>>>>>>>>>>>>
%
   \begin{figure}[h]
   \centering
   \includegraphics[width=80mm]{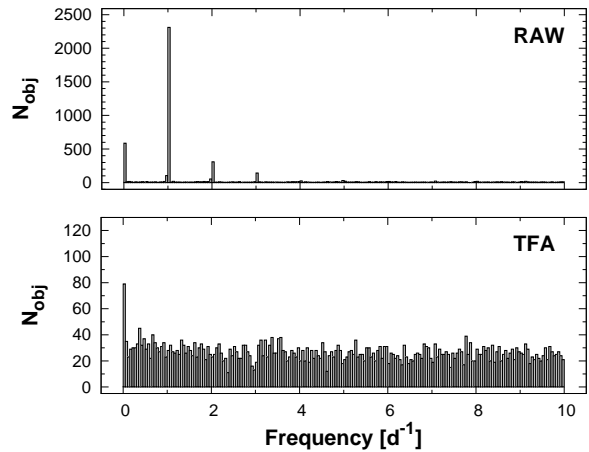}
      \caption{As in Fig.~5, but for field \#79. The TFAd data were 
               obtained with $N_{\rm TFA}=600$.} 
         \label{fig6}
   \end{figure}

The way TFA is currently implemented is not optimal. We select 
a rather large sample of templates that are used for all targets 
in the field. Since we do not know a priori which template members 
are `valuable' we need to select a sufficiently large set in order 
to be more confident that the necessary templates for any given 
target are included. In keeping the `most relevant' templates (or 
the linear combinations of those) some addition selection could be 
incorporated by using principal component decomposition, similarly 
to that of the SysRem algorithm of Tamuz et al. (2005). However, 
applying this or any other similar method, we have to make the 
decision of where to cut the main principal components. It may 
happen that certain systematics (e.g., transients) will escape our 
attention, since they will not enter in the main principal components, 
because the statistics we use (e.g., the eigenvalues) in the cut 
may not be sensitive to the given systematics due to their relatively 
low incidence rates, low amplitudes or short durations. 

In another effort to identify a more economical selection of 
templates, we test if the best fitting single template selected 
from the standard TFA template set gives similar result to the 
algorithm used in the original framework devised by KBN. By using 
the large template sets chosen above in this section, we compute the 
unbiased estimates of the standard deviations (see Appendix A) for 
all stars in the two fields. Then, for each object, we select that 
single-template which yields the lowest $\sigma_{\rm TFA}$ value. 
In Fig.~7 we plot ratio of these standard deviations versus the 
standard deviations computed within the original framework of KBN. 
It is clear that the optimum single-template selection 
{\em always performs less effectively} than the standard multi-template 
method. There is a dependence of the gain of the multiple template 
implementation of TFA as a function of the LC scatter. This effect 
is due to the increase of random noise of the light curves that 
cannot be filtered out by TFA. We also note that although the peak 
frequency distribution for the above optimum single-template fit 
is similar to that of the RAW data, there is a visible decrease 
in the number of stars with integer d$^{-1}$ frequencies. These all 
show that the multi-template implementation of TFA performs 
significantly better than its optimized single-template variant. 

%
%>>>>>>>>>>>>>>>
%  FIGURE 7
%>>>>>>>>>>>>>>>
%
   \begin{figure}[h]
   \centering
   \includegraphics[width=85mm]{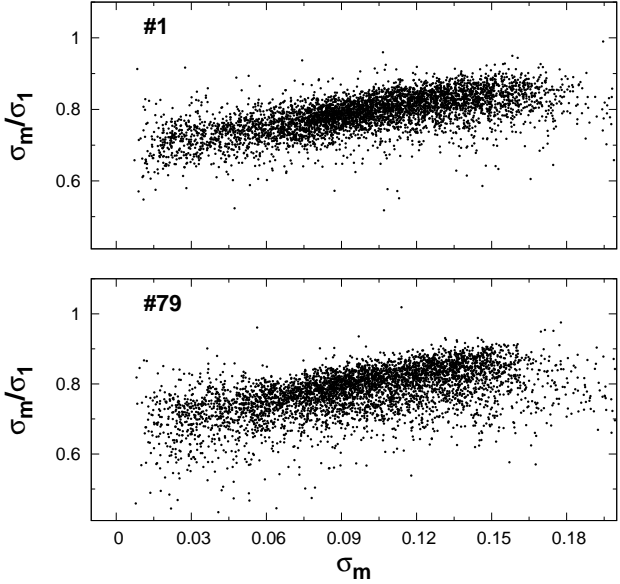}
      \caption{Unbiased estimates of the standard deviations obtained in 
               the multi-template ($\sigma_m$) and optimum single-template 
	       ($\sigma_1$) applications of TFA.} 
         \label{fig7}
   \end{figure}

It is important to test the signal-to-noise ratio (SNR) above which 
one can declare a signal as a significant detection. Although the 
statistical properties of DFT on continuously sampled data are 
well known (e.g., Foster 1996 and references therein), for gapped 
and randomly sampled datasets the situation could be more complicated. 
Therefore, we estimate the confidence level by following the same 
method as in Nagy \& Kov\'acs (2006). Over a thousand time series were
generated by using pure Gaussian white noise on the time base of 
randomly selected objects in each field. Then, these time series 
were treated in the same way as the real observational data and 
TFA/DFT analyses were performed in the $[0.0,10.0]$d$^{-1}$ frequency 
range (with $500$ and $600$ templates in fields \#1 and \#79, 
respectively). After finding the peak frequency, we compute the 
signal-to-noise ratio, defined as follows

% 
%%%%%%%%%%%%%%
%   EQ. (1)
%%%%%%%%%%%%%%
%
\begin{eqnarray}
{\rm SNR} = {A_{\rm peak}-\langle A_{\nu}\rangle\over \sigma_{A_{\nu}}} 
\hskip 2mm .
\end{eqnarray}
Here $A_{\rm peak}$ is the amplitude at the highest peak in the 
spectrum, $\langle A_{\nu}\rangle$ is the average (over the frequency 
values \{$\nu$\} of the spectrum in the passband of the analysis) and 
$\sigma_{A_{\nu}}$ is its standard deviation, computed by an iterative 
$5\sigma$ clipping. We note that, as it follows from the above definition, 
for colored noise, SNR is a function of the frequency range of analysis. 
In the most common situation of decaying low-frequency noise (the 
so-called ``red noise'' -- see, e.g., Pont et al. 2006), by increasing 
the frequency range, SNR also increases. Here we assume that in the 
frequency interval used, the data are well-filtered from systematics 
(corresponding to the main source of red noise) and the flat (white) 
noise spectra generated in the simulations represent the real distribution 
closely enough. 

The distribution functions of the so-obtained SNR values are displayed 
in Fig.~8. It is remarkable that the pattern of the false alarm 
probability (FAP) is very similar for both fields (the small differences 
can be attributed to the finite sample size). Based on this diagram, 
we place our cutoff value at SNR~$=7.0$, corresponding to $\sim 1$\% FAP. 
Although this limit might suggest a rather low rate of false detection, 
in practice the rate of rejected detections is much higher, because 
visual inspection and other criteria (e.g., blend situation) can reveal 
additional signatures of false detection. As a result, from the originally 
selected 450 variables satisfying the SNR~$>7.0$ criterion only slightly 
greater than 80\% survived and entered our variable list. Decreasing the 
SNR cutoff to 6.5 we would select 1015 variables, with an estimated very 
large real false alarm rate, since the formal rate is already $\sim 4$\% 
at this SNR value. On the other hand, a more stringent criterion of 
SNR~$>7.5$ would have led to the loss of some of the interesting 
variables, because this criterion would have been satisfied only by 330 
objects, that is 10\% lower than the number of the finally selected 
variables, based on the DFT analysis.

%>>>>>>>>>>>>>>>
%  FIGURE 8
%>>>>>>>>>>>>>>>
%
   \begin{figure}[h]
   \centering
   \includegraphics[width=85mm]{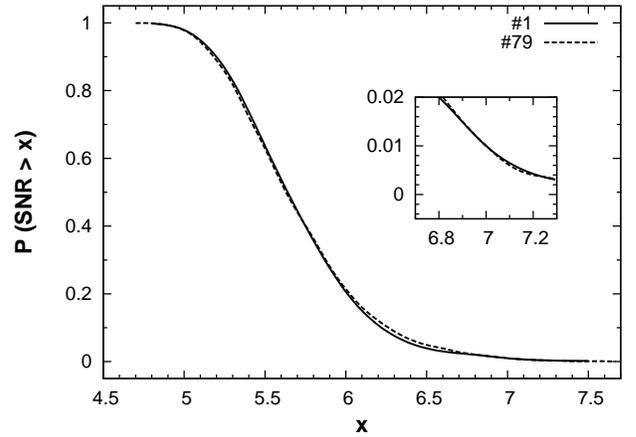}
      \caption{False alarm probability (FAP) diagram for the data 
               used in this paper. Inset shows the close neighborhood 
	       of the cutoff signal-to-noise ratio (SNR) of $7$, 
	       above which FAP is lower than $1$\%.} 
         \label{fig8}
   \end{figure}
%

%%%%%%%%%%%%%%%%%
%  SECTION 4
%%%%%%%%%%%%%%%%%
%
\section{Frequency analysis}
We have performed DFT and BLS frequency analyses on both fields by 
using the optimum template numbers of 500 and 600 for fields \#1 
and \#79, respectively. To avoid faint, noise-dominated stars, 
the template members were selected from stars brighter than 
$V=18.6$~mag. We also performed the same analyses on the RAW data, 
to get an estimate on the effectiveness of TFA. The frequency bands 
of the searches were different for the BLS and DFT runs, since the 
line profiles in the BLS spectra are much narrower that those in 
the more traditional Fourier spectra (see Kov\'acs et al. 2002). 
For the BLS analysis we took the $[0.01,1.0]$~d$^{-1}$ range, 
thereby covering most of the orbital periods of eclipsing binaries, 
the prime targets of the BLS search. For the DFT spectra we have 
a wider range of $[0.0,10.0]$~d$^{-1}$, since we want to cover a 
wide class of variables from long-period red variables to the 
much shorter period $\delta$~Sct stars (by covering at least those 
with reasonably long period to fit in the above frequency range). 
In both cases we used $3\times10^5$ frequency steps.    

Variable star candidates have been selected in three main steps. 
First we selected targets exceeding a given SNR value of their spectra. 
This cutoff value was set to $7.0$ for the DFT spectra and to $\sim9$ 
for the BLS spectra (this latter value was approximately selected on 
the basis of our earlier experiments that showed the BLS spectra to 
be more vulnerable to false alarms). In this way we selected some 
200 stars from each of the two fields (many of the variables were 
common among the DFT and BLS candidates). Then, all variables were 
visually inspected and further filtered for possible false alarms. 
In the course of this we also assigned some preliminary classification 
to the variables. In the final step the independent classifications 
of the authors have been discussed and the final list of variable star
candidates was selected. We have also performed some supplementary 
analyses. For example, we compared in more detail the long-period 
variables in the RAW and in the TFA-filtered time series and investigated 
the effect of target-specified time base on the signal detection (an 
effect that could be important in the case of targets with number of 
data points lower than the average). We show the coordinates and the 
most important parameters of the so-obtained variables in 
Tables~1 and 2\footnote{Time series of the objects given in Tables~1 
and 2 are accessible at {\em http://www.konkoly.hu/staff/kovacs}}.  

We caution that our classification is {\em preliminary} and very 
{\em approximate}. Except for the trivial cases (e.g., typical 
fundamental mode RR~Lyrae stars), variable star classification 
(i.e., the first step toward the estimation of the physical 
parameters) is a very difficult task 
in heterogeneous stellar systems. In our simple classification 
scheme we considered: (i) the period; (ii) the approximate 
brightness in $V$ and color $V-R$; (ii) the light curve shape. 
These have led to 11 classes, with the wide group of long-period 
variables (LPVs) including objects on the AGB and beyond and, of 
course, miscellaneous variables, the ones that show definite 
variability but indefinite classification based on the parameters 
listed above. The broad definitions of the variable classes are 
as follows.

\noindent\parbox[t]{15mm}{\bf LPV    :} \parbox[t]{70mm} 
{not eclipsing and period longer than $\sim10$ days and 
brighter than the HB level of $V\sim19$~mag}

\noindent\parbox[t]{15mm}{\bf FU Cep :} \parbox[t]{70mm} 
{fundamental mode Cepheid}

\noindent\parbox[t]{15mm}{\bf FO Cep :} \parbox[t]{70mm} 
{first overtone mode Cepheid}

\noindent\parbox[t]{15mm}{\bf SO Cep :} \parbox[t]{70mm} 
{second overtone mode Cepheid}

\noindent\parbox[t]{15mm}{\bf RRab   :} \parbox[t]{70mm} 
{fundamental mode RR~Lyrae}

\noindent\parbox[t]{15mm}{\bf RRc    :} \parbox[t]{70mm} 
{first overtone mode RR~Lyrae}

\noindent\parbox[t]{15mm}{\bf EB     :} \parbox[t]{70mm} 
{eclipsing binaries of any type}

\noindent\parbox[t]{15mm}{\bf B      :} \parbox[t]{70mm} 
{$\sim$B-type pulsators, $V\lesssim18.0$, $V-R\lesssim0.1$}

\noindent\parbox[t]{15mm}{\bf $\delta$~Sct :} \parbox[t]{70mm} 
{frequency is greater than $\sim5$d$^{-1}$ and $0\lesssim V-R\lesssim0.3$}

\noindent\parbox[t]{15mm}{\bf Hump   :} \parbox[t]{70mm} 
{hump/eruption in the light curve}

\noindent\parbox[t]{15mm}{\bf Misc   :} \parbox[t]{70mm} 
{anything that cannot be classified}

\noindent
The following comments apply to the variable star inventories provided
in Tables~1 and 2.   

\begin{itemize}
\item
Magnitudes have been derived by using Eq.~(1) of Alcock et al. (1997). 
We note that these magnitudes are approximate and can be used only as 
a rough guide for variable classification. 
\item
Variables without a dominant frequency component (leading to large 
-- above the noise level -- SNR values) will not get promoted into
our lists.
\item
Blends are defined as those stars that have close positions and periods 
to those of the target but they appear to have lower SNR values (and 
concomitantly lower amplitudes). Except for a few cases, the distinction 
between the blend and the source was unique. These stars are commented 
together with other peculiarities in Appendix B. Blends are not
reported in Tables~1 and 2.
\item
Frequencies can be ambiguous within integer fractions for EB stars.
\item
There are some near integer d$^{-1}$ variables that may well be LPVs, 
but it is impossible to decide it from the present data. In particular, 
some of the red, short-periodic Misc variables might actually be LPVs.
\item
The slowly varying LPVs -- those that change their brightness 
over the time scale of the observational time span -- may be subject 
to ``signal killing''. That is, because for long-period targets it is 
easier to find templates with similar long-term changes, it is also 
easier to subtract the signal by the proper combination of these 
templates. However, we checked the original (non-TFAd) LCs of all 
objects, and we found only a handful of them disappeared after 
TFA-filtering. Further examination of these cases revealed that they 
all had rather large scatter and variations reminiscent of systematics. 
Therefore, we declared them as such.  
\item
There are some stars classified as RRab?, because the period, color 
and brightness are in the appropriate ranges. Otherwise the shape 
of the LC is not of a typical RRab and the amplitude is rather low. 
Similar note holds also for other types of variables.  
\item
Although the prime selection was made on the basis of low false alarm 
probability, there is still some chance that a small fraction 
(maybe 5--10\%) of the variables with low SNR are, in fact, false 
alarms.    
\item
TFA detections are those with one of the following properties: 
(i) the RAW data show typical systematics at integer d$^{-1}$ peak 
frequencies that disappear in the TFA spectra with the concomitant 
appearance of significant peak(s) at non-integer d$^{-1}$ frequencies; 
(ii) as in (i) but the RAW frequencies are at non-integer d$^{-1}$ 
frequencies. We also note that when the SNR of the TFA spectrum was 
higher but the RAW spectrum gave the same frequency at a lower SNR 
level (but above the reasonable detection limit) we considered this 
as a non-TFA detection.

\end{itemize}

%
%%%%%%%%%%%%%%%%
%   TABLE 1
%%%%%%%%%%%%%%%%                                                    
%
\begin{table*}[h]
\caption[]{Variable stars in the MACHO field \# 1.}
\begin{flushleft}
\begin{tabular}{ccclrrc}
\hline\hline
 ID  &  V  & V$-$R  &  f[$d^{-1}$] &  SNR  &  Type  & TFA \\
\hline
05011976$-$6921083 & 16.889 &  0.101 &  1.80367  &  14.5 & B       & 1\\
05011992$-$6919188 & 17.585 &  0.049 &  0.37573  &  10.8 & B       & 0\\
05011998$-$6917495 & 18.066 &  0.430 &  0.252205 &  16.9 & EB      & 0\\
05012021$-$6920315 & 18.940 &  0.582 &  7.82853  &   7.0 & d Sct?  & 1\\
05012029$-$6920248 & 19.615 &  0.113 &  0.02328  &  13.9 & LPV     & 0\\
\hline
\end{tabular}                                                         
\end{flushleft}                                                       
{\footnotesize\underline {Notes:}

$-$ The object identification follows the notation of the 2MASS catalog, 
    namely it is simply the abbreviated J2000 (RA,Dec) coordinates of the 
    object. For example, 05011976$-$6921083 corresponds to RA=05:01:19.76, 
    Dec=$-$69:21:08.3.  

$-$ The frequencies for type EB variables are given with larger number 
    of digits, because of the higher sensitivity of the eclipse shape to
    the precision of the orbital period.

$-$ The definition of the signal-to-noise ratio (SNR) and classification 
    scheme are described in the text.

$-$ Single digits under the column ``TFA'' mean exclusive TFA detections 
    if ``1'' and pre-TFA detections if ``0''.    

$-$ The excerpt shown is a part of the full table which is available 
    only in the electronic version of this paper.
}                                          
\end{table*}

%
%%%%%%%%%%%%%%%%
%   TABLE 2
%%%%%%%%%%%%%%%%                                                    
%
\begin{table*}[h]
\caption[]{Variable stars in the MACHO field \# 79.}
\begin{flushleft}
\begin{tabular}{ccclrrc}
\hline\hline
 ID  &  V  & V$-$R  &  f[$d^{-1}$] &  SNR  &  Type  & TFA \\
\hline
05122727$-$6920387 & 16.948 &  0.907 &  0.04885  &   8.9 & LPV     & 1\\
05122740$-$6919308 & 16.270 &  0.091 &  1.31131  &  11.0 & B       & 1\\
05122742$-$6903459 & 18.708 &  0.552 &  7.12253  &   7.2 & d Sct   & 1\\
05122749$-$6919426 & 18.831 &  0.424 &  2.76320  &  26.5 & RRab    & 0\\
05122754$-$6918475 & 18.626 &  0.046 &  0.47837  &   7.8 & Misc    & 0\\
\hline
\end{tabular}                                                         
\end{flushleft}                                                       
{\footnotesize\underline {Note:}

$-$ See notes to Table~1. 
}                                          
\end{table*}

In Table~3 we provide the statistics of the variables detected in 
the two fields. The large number of {\em exclusive} TFA detections 
is striking (these are the signals that did not hit the detection 
limit in the RAW data). Interestingly, the relative number of TFA 
detections in both fields is nearly the same. By applying TFA, we 
have {\em doubled} the number of detected variable stars. The 
variability rate (number of variables versus total number of stars 
analyzed) is 3--5\% in the two fields, very similar to the recent 
rate derived from the HATNet database on selected Galactic fields 
(Kov\'acs \& Bakos 2008). It is also important to mention that TFA 
is useful even if the signal is detected already in the RAW data. 
If systematics are present, then signal reconstruction with the aid 
of TFA helps to derive more accurate signal, and eventually, more 
accurate stellar parameters.

%
%%%%%%%%%%%%%%%%
%   TABLE 3
%%%%%%%%%%%%%%%%                                                    
%
\begin{table}[h]                                                     
\caption[]{Variables in the sample of fields \#1 and \#79}                                                    
\begin{flushleft}                                                     
\begin{tabular}{lrrrr}                                         
\hline
Type & 
\multicolumn{2} {c} {\#1} &  
\multicolumn{2} {c} {\#79} \\
\multicolumn{1} {c} {} &
\multicolumn{1} {r} {$ND_{\rm RAW}$} &
\multicolumn{1} {r} {$ND_{\rm TFA}$} &
\multicolumn{1} {r} {$ND_{\rm RAW}$} &
\multicolumn{1} {r} {$ND_{\rm TFA}$} \\
\hline\hline
 LPV          & 18  &  24  &  34 &  32 \\  
 FU Cep       &  1  &   1  &   3 &   0 \\  
 FO Cep       &  3  &   1  &   0 &   1 \\  
 SO Cep       &  1  &   1  &   1 &   0 \\  
 RRab         & 10  &   4  &  16 &   3 \\  
 RRc          &  3  &   3  &   1 &   1 \\     
 EB           & 27  &   7  &  30 &  21 \\  
 B            & 11  &   8  &  12 &  14 \\   
 $\delta$~Sct &  1  &   9  &   4 &   5 \\     
 Hump         &  1  &   0  &   1 &   0 \\  
 Misc         &  9  &  23  &  17 &  25 \\  
\hline                                                                
 Sum          & 85  &  81  & 119 & 102 \\
\hline                                                                
\end{tabular}                                                         
\end{flushleft}
{\footnotesize\underline {Notes:}

  $ND_{\rm RAW}$ = number of detections in the original (RAW) data 
  
  $ND_{\rm TFA}$ = number of {\em exclusive} detections due to TFA-filtering 
}                                       
\end{table}

To exhibit the signal detection and reconstruction power of TFA, we 
show three examples. In Fig.~9 we plot the DFT frequency spectra of 
05011976-6921083 from field \#1, a short-periodic sinusoidal variable, 
classified as type B (blue, relatively bright, presumably pulsating 
variable). The total (peak-to-peak) amplitude is only 0.03~mag in 
MACHO instrumental ``b'' passband. In the spectrum of the RAW 
data, the true signal frequency is hidden in the 1d$^{-1}$~side lobe 
of the systematics. This component clearly stands out in the spectrum 
after applying TFA. Reconstruction of the signal with this period results 
in a substantial improvement in the noise level and in the shape of 
the signal (on which no assumption has been made during the reconstruction). 
However, it is important to note that the reconstructed signal displays 
the true signal with the residuals computed by the full signal model 
of signal$+$trend$+$noise and the latter component is simply the 
residual after subtracting the signal and trend components from the 
RAW data. Because the procedure involves the least squares fit of a 
considerable number of parameters, the residuals will be {\em biased}, 
i.e., they show smaller scatter as they should, due to the extra 
correlation generated among them by the above fit. Due to this effect 
(mentioned already in Sect.~3.2 and discussed in detail in Appendix 
A), the true noise level is some 30\% higher than it is shown in the 
figure. Nevertheless, even with this caveat, the application of TFA 
clearly leads to a cleaner signal shape (which, in turn, might suggest
a different classification -- e.g., that of EB -- but the data are 
insufficient to be conclusive in this case.)

%
%>>>>>>>>>>>>>>>
%  FIGURE 9
%>>>>>>>>>>>>>>>
%
   \begin{figure}[h]
   \centering
   \includegraphics[width=70mm]{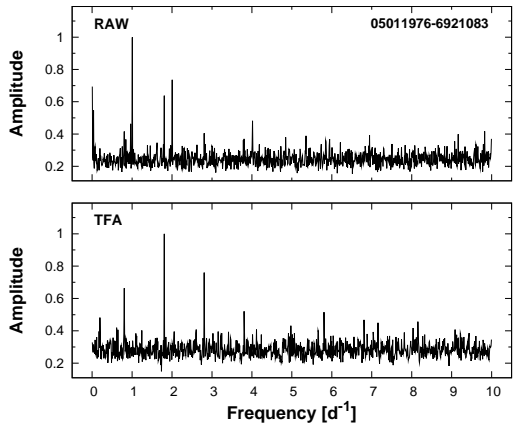}
      \caption{Example of the detection of a signal strongly dominated  
               by systematics (see the high peaks at integer d$^{-1}$ 
	       frequencies in the RAW spectrum). The variable is from 
	       field \#1. } 
         \label{fig9}
   \end{figure}
%

%
%>>>>>>>>>>>>>>>
%  FIGURE 10
%>>>>>>>>>>>>>>>
%
   \begin{figure}[h]
   \centering
   \includegraphics[width=70mm]{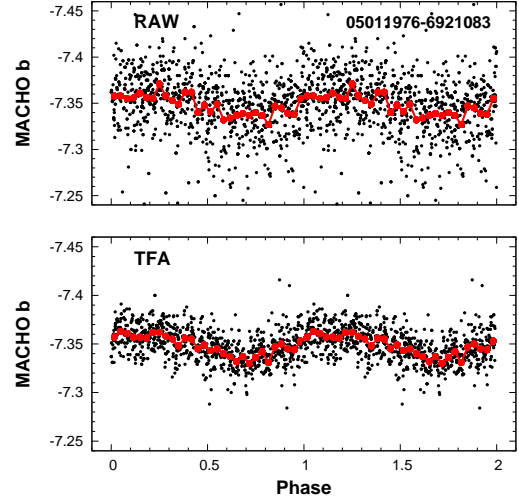}
      \caption{TFA reconstruction of the B-type variable shown in Fig.~9. 
               Please check the note in the text on the scatter of the 
	       TFA-reconstructed light curve.} 
         \label{fig10}
   \end{figure}
%

%
%>>>>>>>>>>>>>>>
%  FIGURE 11
%>>>>>>>>>>>>>>>
%
   \begin{figure}[h]
   \centering
   \includegraphics[width=70mm]{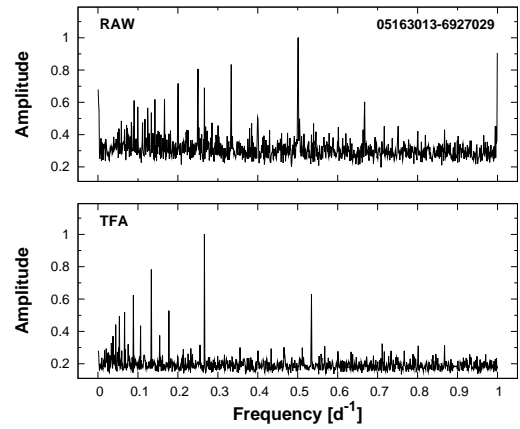}
      \caption{Example of the detection of a signal nearly completely 
               smeared by systematics in the original (RAW) data. The 
	       frequency spectra were computed by the BLS algorithm. 
	       The variable is from field \#79.} 
         \label{fig11}
   \end{figure}

Figures 11 and 12 show a similarly strong TFA detection for the 
binary star 05163013-6927029 from field \#79, discovered with much 
higher SNR by BLS than by DFT frequency analysis. The BLS frequency 
spectrum of the RAW data clearly shows the peak structure at 
$(0.5/n)$d$^{-1}$, reminiscence of the daily systematics present 
in the BLS spectra of most of the light curves. The true signal 
component shows up as a middle-size peak in the forest of the 
subharmonics of the daily trend. We note that the confusion in the 
case of BLS spectra is exaggerated due to the (sub)harmonic structure 
inherent to the BLS method (or to any other method, based on 
period-folding -- see Kov\'acs et al. 2002) . The reconstructed 
and folded LC shown in the lower panel of Fig.~12 clearly exhibits 
a shallow secondary eclipse, not even suspected in the RAW LC. 
The bias factor for the residual scatter is the same here as 
mentioned in the context of Fig.~10, since for field \#79 we have 
more data points, but, at the same time, we also have more extended 
TFA templates. 

%
%>>>>>>>>>>>>>>>
%  FIGURE 12
%>>>>>>>>>>>>>>>
%
   \begin{figure}[h]
   \centering
   \includegraphics[width=70mm]{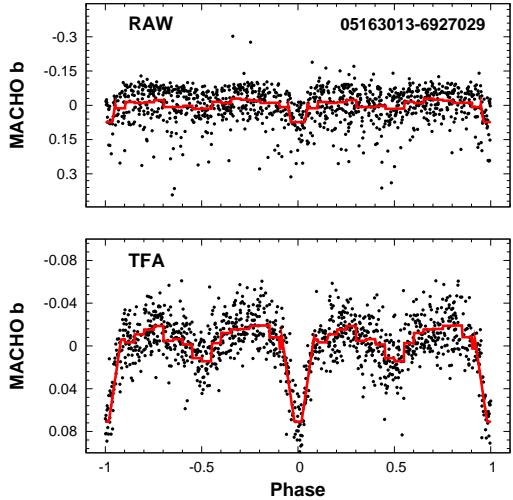}
      \caption{TFA reconstruction of the EB variable shown in Fig.~11.} 
         \label{fig12}
   \end{figure}

Our third example is devoted to demonstrate that systematics may 
not necessarily be exhibited in the Fourier domain. As mentioned 
in Sect.~3.2, transients may not always be properly subtracted 
from the images. Since these phenomena are non-periodic, they 
usually do not affect the frequency spectra in a significant way. 
Figure 13 shows the DFT spectra of such an object (an RR~Lyrae 
star). The signal is easily detected already in the RAW data. 
Even more, we see a decrease in the SNR in the TFA spectrum. 
This is due to the fact that the observed signal is strong 
(relative to the systematics) and in the signal search mode 
TFA assumes a zero target function. When this assumption is 
lifted, i.e. a full model is used and a signal reconstruction 
is made, we obtain the result shown in Fig.~14. Nearly all 
outlying data points are successfully corrected, with a slight 
accompanying change in the LC shape. 

%
%>>>>>>>>>>>>>>>
%  FIGURE 13
%>>>>>>>>>>>>>>>
%
   \begin{figure}[h]
   \centering
   \includegraphics[width=70mm]{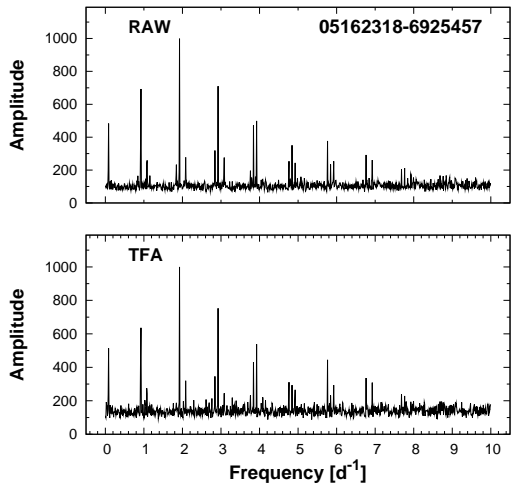}
      \caption{Example of the detection of a DFT signal in the 
               RAW data. The particular systematics present in 
	       this object do not affect the DFT frequency spectra.      
	       The lower frequency sidelobes visible at the alias 
	       components are due to the first harmonic (and the 
	       aliases) of the fundamental frequency. The variable 
	       is from field \#79.} 
         \label{fig11}
   \end{figure}
%

%
%>>>>>>>>>>>>>>>
%  FIGURE 14
%>>>>>>>>>>>>>>>
%
   \begin{figure}[h]
   \centering
   \includegraphics[width=70mm]{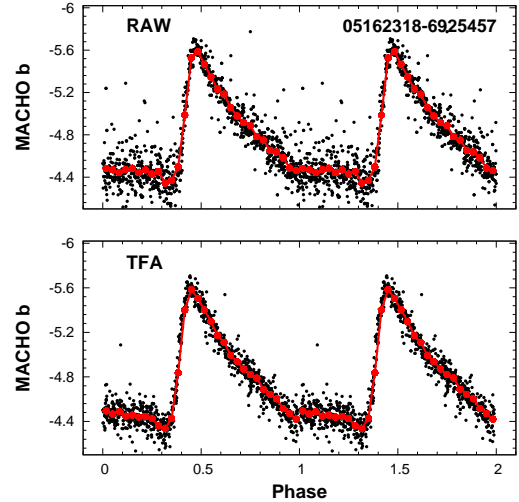}
      \caption{TFA reconstruction of the RR~Lyrae variable shown in Fig.~13. 
               Systematics are exhibited as outlying data points, causing 
	       only a small effect on the frequency spectrum.} 
         \label{fig14}
   \end{figure}

For a rough location of the variables in the physical parameter 
space, in Fig.~15 we overplot the detected variables on the CMD 
of \#79. We see that the variables basically follow the distribution 
of the constant stars, with the exception of the luminous red 
variables and perhaps with some mild surplus of red clump variables 
centered at $V=19.2$ and $V-R=0.6$. The lack of variables on the 
giant branch between $V=18$ and $17$~mag is very striking. As 
expected from their short evolutionary life times, we do not have 
that many post AGB stars either. Although we have a reasonably large 
number of RR~Lyrae variables (41 altogether), they do not seem to form 
a well-defined separate group at the horizontal branch between $V=19$ 
and $20$~mag, but they are rather intermingled with other types of 
variables. On the other hand, the main sequence is nearly uniformly 
populated, from the the bright end down to the faintest stars in our 
sample. 

%
%>>>>>>>>>>>>>>>
%  FIGURE 15
%>>>>>>>>>>>>>>>
%
   \begin{figure}[h]
   \centering
   \includegraphics[width=70mm]{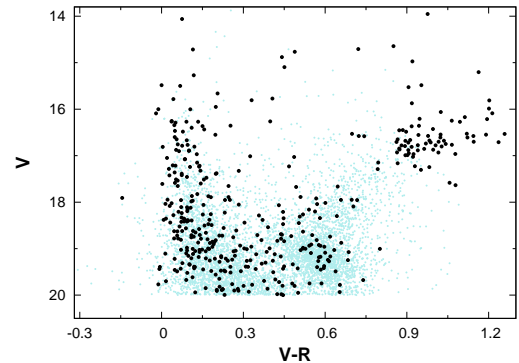}
      \caption{Color-magnitude diagram of field \#79. Overplotted 
               are the 387 variables detected in fields \#1 and 79.} 
         \label{fig15}
   \end{figure}
%

%%%%%%%%%%%%%%%%%
%  SECTION 5
%%%%%%%%%%%%%%%%%
%
\section{Discussion and Conclusions}
We employed the Trend Filtering Algorithm (TFA) combined with Fourier 
(DFT) and box-search (BLS) frequency analyses to explore the variable 
star content of a subset of the MACHO Project database for the Large 
Magellanic Cloud. From direct analyses, a large part of the database 
yields too many false signals due to the strong systematics present 
in the data. We found that about 50\% of detectable variables were not 
detected by standard strategies in the MACHO Project database for the 
samples we investigated ($\sim 5300$ objects in each of the fields 
\#1 and \#79, down to $V=20.0$~mag). About 60\% of all stars are 
dominated by trends.    

Due to the relatively small number of data points per object, we 
investigated the optimum number of TFA templates to be used in 
the survey. In an ideal case the representative model of a time 
series is reached when the increase of the number of fitting functions 
does not lead to a further decrease of the unbiased estimate of the 
standard deviation of the residuals between the input time series 
and the model. Unfortunately, for the MACHO Project time series, it 
seems that the required number of templates exceeds half of the number 
of data points. This would lead to a prohibitive overuse of the TFA 
fit and, as a result, to a significant increase in false alarms. 
Fortunately, the distribution of the peak frequencies of the Fourier 
spectra of the TFA-filtered time series -- even with sub-optimal 
template numbers -- shows that the most significant systematics 
exhibited as integer fractions of the diurnal periodicity disappear. 
We also examined the case of very long-period variables (showing 
basically a linear trend), where TFA is more likely to distort the 
light variation and lead eventually to the reduction of the 
true photometric variation. Although we found that the variations 
were real in a few cases and TFA distorted the signal considerably, 
visual inspections of the original (non-filtered) light curves led 
to the conclusion that, in general, the signals were most likely 
long-term systematics.    

We found altogether 387 variables from the total of $\sim10600$ objects 
analyzed in the subsets of the two fields. This variability ratio is 
very close to the value recently derived on a subsample of the HATNet 
database (Kov\'acs \& Bakos 2008). This coincidence is most probably 
accidental, since HATNet surveys the Galactic field stars and detects 
also many variables in the millimagnitude amplitude regime that, for 
the majority of objects, is not accessible by the MACHO Project survey. 
Nevertheless, it is remarkable that with TFA, we are able to detect 
many low-amplitude variables in the few-times 10~mmag regime also in 
the MACHO Project database. We think that most of these variables are 
indeed ``normal''-amplitude variables, but due to significant crowding 
and blending they appear as low-amplitude variables.
 
We find that the application of TFA in variable search increases 
our ability to find many low-amplitude variables that might be not 
accessible in the raw photometric data. This may lead to exciting 
new studies on various fields related to variable stars. Here we 
mention only three topics of immediate interest. 
\begin{itemize}
\item
Low-amplitude pulsating variables across the HR-diagram: There 
are of course interesting questions to raise everywhere. For example, 
excitation map of B-type stars is not well known, because of the 
delicate dependence on the metal content (Karoff et al. 2008). 
Limits of variability in the $\delta$~Scuti regime is still to be 
determined. This would be especially interesting, since linear pulsation 
theory predicts excitations throughout this part of the HR-diagram 
(e.g., Breger, Lenz, \& Pamyatnykh 2008). There are regions above 
the Horizontal Branch (HB) still to explore: the red-clump stars, the 
high-metallicity, low-temperature `cousins' of the HB stars; variables 
on the giant branch and in the AGB phase. The change of role of 
convection in the context of the classical $\kappa$-mechanism as we 
reach higher luminosity levels is not understood. The relatively 
well-confined instability strip of classical pulsators becomes very 
poorly defined at higher luminosities. Although recent investigations 
explored many interesting properties of the red giant variables 
(e.g., Soszy\'nski et al. 2007 and references therein), they also 
revealed their complexity and mixed physical nature.  
\item
Ultra low-amplitude (ULA) classical pulsators: Although the physical 
mechanism in establishing the borders of the instability strip 
of Cepheids and RR~Lyrae stars is more or less known, it is still 
unclear how the pulsation stops and if other forms of pulsation  
at low amplitudes could exist outside the instability strip. From a 
subsample of the MACHO LMC Cepheids Buchler et al. (2005) found 14 
objects with characteristic amplitudes of $0.005$--$0.01$~mag. 
Interestingly, these stars apparently follow the PLC relation spanned 
by the `normal' large-amplitude Cepheids. The above sample of ULA 
Cepheids has recently been extended by Soszy\'nski et al. (2008). 
Some theoretical considerations and numerical simulations (Buchler \& 
Koll\'ath 2002; Bono et al. 1995) suggest that in the case of `soft' 
bifurcations (when the pulsation growth rates are in the same order 
as the evolutionary rates), the transition from various pulsation 
states may last several thousand years, so we may have rare, but 
observable events. 
\item
Shallow-eclipsing binaries: These are important objects in extending 
the determination of stellar parameters toward the very low-mass 
($<0.1$~M$_{\odot}$ regime. Only a handful of objects span the low-mass 
tail of the mass--radius relation, so discovering additional objects 
would be very important in clarifying the source of the systematic 
difference between the observations and the currently available 
theory (Beatty et al 2007). Depending on the configuration, the depth 
of the eclipse can be very low, but a typical M star with an F primary 
yields a $\sim0.01$~mag deep eclipse. It is clear that detection of 
such events (even if it is periodic) requires good quality data avoiding 
contamination with systematic effects. 
\end{itemize}

Although the discovery of low-amplitude variables is a very important 
task, it is also a very hard one. We not only have a faint signal but 
it is also intermingled with the effect of colored noise due to stellar 
activity and instrumental systematics as discussed above. Furthermore, 
on the top of these, there is the blending issue. This problem has 
recently been highlighted by the search for extrasolar planets by transit 
methods (Brown \& Latham 2008) and problems concerning the measured PL 
relations in distant galaxies (Vilardell, Jordi \& Ribas 2007). The 
problem is quite common in all photometric works (Kiss \& Bedding 2005) 
and can be tackled at a certain degree by various data reduction methods 
(i.e., image subtraction and profile analysis, see Hoekstra, Wu \& Udalski 
2005) but in tough cases we need combined efforts by experts on large 
telescope photometry, spectroscopy and stellar modeling (Mandushev et al. 
2005).

%%%%%%%%%%%%%%%%%%%%
% ACKNOWLEDGEMENTS
%%%%%%%%%%%%%%%%%%%%
%
\begin{acknowledgements}
This paper utilizes public domain data obtained by the MACHO Project, 
jointly funded by the US Department of Energy through the University 
of California, Lawrence Livermore National Laboratory under contract 
No. W-7405-Eng-48, by the National Science Foundation through the 
Center for Particle Astrophysics of the University of California under 
cooperative agreement AST-8809616, and by the Mount Stromlo and Siding 
Spring Observatory, part of the Australian National University. G.~K. 
thanks for the support of the Hungarian Scientific Research Fund (OTKA, 
grant No. K-60750). We are grateful to the National Information 
Infrastructure Development (NIIF) Program for providing CPU time for 
most of the computations presented in this paper (project No. 1109). 
DLW acknowledges the support in the form of a Discovery Grant by the
Natural Sciences and Engineering Research Council of Canada (NSERC).

\end{acknowledgements}

%
%%%%%%%%%%%%%%%%%
%  APPENDIX A
%%%%%%%%%%%%%%%%%
%
\begin{appendix} 
\section{Unbiased estimation of the standard deviation of the residuals 
of least squares fits to sparsely sampled target functions}
Here we derive a formula for the unbiased estimate of the standard 
deviation of the residuals when the target function is sparsely 
sampled relatively to the fitting functions of a Least Squares (LS) 
problem. This situation may occur in the case of the application 
of TFA, when we filter with a template set defined on a more extended 
time base than that of the target (a situation, that could be quite 
common in photometric databases containing data of diverse quality). 

The LS problem is defined in the standard way

% 
%%%%%%%%%%%%%%
%   EQ. (A1)
%%%%%%%%%%%%%%
%
\begin{eqnarray}
s^2 & = & 
{1\over n}\sum_{i=1}^{n} \big[y(i)-\sum_{j=1}^{m}a_jx_j(i)\big]^2 = min. \hskip 2mm ,
\end{eqnarray}
where $\{y(i)\}$ is the target function (in the following: time series), 
$\{x_j(i); j=1,2,...,m\}$ are the fitting functions with the corresponding 
regression coefficients $\{a_j\}$. We note that the fitting functions  
might include any functions that contribute to the signal; for example, 
they might contain Fourier components and TFA templates. At the minimum 
variance, $s$ stands for the root mean squares (RMS) of the residuals. 
Our assumptions are the following: 
\begin{itemize}
\item
All $\{x_j(i)\}$ are sampled in the same time base $\{t(i)\}$. 
\item 
The sampling time base $\cal T$ of $\{y(i)\}$ is a nonzero subset 
of $\{t(i)\}$. This part of $\{y(i)\}$ contains $n_1$ data points. 
\item 
$\{y(i)\}$ is assumed to be zero-averaged by supplying ``zeros'' at 
the moments where it is not defined (in respect to $\{t(i)\}$). 
\item
The noise $\{\eta(i)\}$ is additive in $\{y(i)\}$ with the following 
expectation values: $E(\eta(i))=0$, $E(\eta(i)\eta(j))=\delta_{ij}\sigma^2$. 
\item
The noiseless signal can be represented by the assumed model given 
by the best-fitting linear combination of $\{x_j(i)\}$. 
\end{itemize}

With the above assumptions, the target time series reads as follows

% 
%%%%%%%%%%%%%%
%   EQ. ...
%%%%%%%%%%%%%%
%
\[ y(i) = \left\{
\begin{array}{ll}
0                               & \mbox{if $t(i)\not\in\cal T$ } \\
y_0(i)+\eta(i)-\langle y_0(i)+\eta(i)\rangle & \mbox{if $t(i)\in\cal T$ }
\end{array}
\right. \] 
Here 
$\langle y_0(i)+\eta(i)\rangle={1\over n_1}\sum_{i\in\cal T}[y_0(i)+\eta(i)]$,  
$\{y_0(i)\}$ is the noiseless part of the target function. With 
$g_{kl}=\sum_{i=1}^n x_k(i)x_l(i)$ and $b_{k}=\sum_{i=1}^n x_k(i)y(i)$ 
the solution of the LS problem reads as 
% 
%%%%%%%%%%%%%%
%   EQ. (A2)
%%%%%%%%%%%%%%
%
\begin{eqnarray}
a_j & = & \sum_{i=1}^m h_{ij}b_i \hskip 2mm ,
\end{eqnarray}
where $\{h_{ij}\}$ is the inverse of $\{g_{ij}\}$ 
($\sum_{j=1}^m h_{ij}g_{jk}=\delta_{ik}$). With these, the expectation 
value of the square of RMS can be written in the following form 
% 
%%%%%%%%%%%%%%
%   EQ. (A3)
%%%%%%%%%%%%%%
%
\begin{eqnarray}
E(s^2) & = & 
{1\over n}\sum_{i=1}^{n} y^2(i)-\sum_{j,k=1}^{m}h_{jk}b_jb_k \hskip 2mm .
\end{eqnarray}
To compute the unbiased estimate of $\sigma^2$, we need to evaluate the 
expectation value of $s^2$. This is a straightforward computation, but 
we need to consider that the lack of data points in $\{y(i)\}$ as 
given above, will introduce extra correlation. Finally, we end up with 
the following expression
% 
%%%%%%%%%%%%%%
%   EQ. (A4)
%%%%%%%%%%%%%%
%
\begin{eqnarray}
n{E(s^2)\over\sigma^2} & = & n_1 - 1    
- \sum_{j,k=1}^{m} h_{jk}(C_{jk}-S_jS_k) \hskip 2mm , 
\end{eqnarray}
where $C_{jk}=\sum_{i\in\cal T} x_j(i)x_k(i)$ and 
$S_j=n_1^{-{1\over 2}}\sum_{i\in\cal T} x_j(i)$. It is easy to see that 
the above expression reduces to the familiar formula for the 
unbiased estimate of the variance if all items of the target 
function have matching time values with those of the fitting 
functions ($C_{jk}=g_{jk}$) and if the latter are also zero-averaged 
($S_j=0$).  

\end{appendix}

%
%%%%%%%%%%%%%%%%%
%  APPENDIX B
%%%%%%%%%%%%%%%%%
%
\begin{appendix} 
\section{Comments to the variable selection of Sect.~4}

\noindent
NOTES ON \#1:
\begin{itemize} 
\item
05021002-6850573 (SNR=33.1, Typ=RRc?) is a close  
companion of 05021018-6850566 (SNR=25.9, Typ=B?). 
Inspection of the LCs does not support an obvious 
blend scenario although their proximity strongly suggests 
that this is the case. 
\item
05022450-6850457  has features that may indicate 
that this is a $\sim 9$~d$^{-1}$ (!) EB
\item
There are stars such as 05022832-6853097 with 
SNR=7.7 that have been omitted, because their 
LCs were ragged. In this particular case the 
alias components at $\sim 1$~d$^{-1}$ are similarly 
strong, but we did not explore the possibility that this 
was the true frequency.
\item
Some of the red, short-periodic Misc candidates might 
actually be LPVs, but we cannot make a distinction, 
due to alias problems. For example, 05055618-6844166 was 
earlier classified as "Misc", but: (i) the difference 
between the heights of the long- and the short-periodic 
peaks is less than 5\%, (ii) the long-periodic LC indeed shows 
long-term features, (iii) the color is very red. So it is 
re-classified as "LPV". The frequency and SNR given in Table~1 
are those of the RAW analysis. Also, 05014177-6920207 has a 
frequency of nearly $\sim 1$~d$^{-1}$. The frequency close to 
zero is also viable. The color is very red. Chances are that 
this is an LPV. 
\end{itemize}

\noindent
NOTES ON \#79:
\begin{itemize}
\item
Stars 05122749-6919426 and 05122739-6919434 are blends of each other. 
Since 05122749-6919426 has a slightly greater SNR, we left this object 
as a variable in the database. However, this choice is not well justified. 
Only further observations with high resolution images can decide which 
one of these stars is the real source of the variation. SuperMACHO Project
images go deeper and have better image quality and could resolve this
ambiguity.
\item
05123435-6904427: 
TFA yields a bit higher power at $1$~d$^{-1}$alias. After visual 
inspection of the unfolded time series, we classified this 
variable as LPV. 
\item
05125486-6918537: 
The RAW light curve is noisy and ragged with a long shallow dimming 
in the first half of the observational time span. After TFA, the 
signal characteristics of SNR=7.3 and $f_{\rm peak}=1.00329$~d$^{-1}$ 
did not seem to be convincing enough, to consider the variability 
as real.
\item
05130892-6919317: Both RAW and TFA yield long periods, but it 
is actually aperiodic within the observational time span. Therefore, 
we assigned the RAW period and SNR to this star. 
\item
05161575-6926342: 
Some part of the long-term variation has been suppressed by TFA.
\item
05162206-6925526: 
There is a relatively strong signal deformation by TFA but the LPV 
classification is secure.
\item
05162318-6925457: is a heavily blended RRab (the three close companions 
are 05162314-6925413, 05162251-6925440 and 05162257-6925491).
\item
05162865-6926167: 
TFA eliminates both the long-term drift and the short-term variation 
(approximately on a time scale of year with amplitudes of 0.05-0.1~mag). 
It is suspected that these variations are related to some systematics.
\item
05162730-6926236: LPV (non-periodic). TFA yields a slightly higher peak 
at $\sim1$d$^{-1}$, that could lead to a misclassification. Frequency, 
SNR entries are those of the RAW data.   
\item
05163808-6926589: TFA/BLS yields SNR=7.7, therefore, due to the SNR 
cutoff of $\sim9$ employed on the BLS spectra, this star does not 
enter in our prime selection. Visual inspection of the LC folded by 
the BLS peak frequency of $0.5746619$d$^{-1}$ might suggest an EB 
classification.
\end{itemize}

\end{appendix}

%%%%%%%%%%%%%%%
% OBJECT LIST
%%%%%%%%%%%%%%%

%%%\listofobjects

\end{document}